\documentclass[12pt]{article}

\usepackage[english,polish]{babel}

\usepackage{pdfpages}
\usepackage{pgf}
\usepackage{graphicx}

\usepackage{times}
\usepackage[T1]{fontenc}

\begin{document}

\begin{center}
{\large EHRENFEST'S THEOREM REVISITED\footnote{Article based on  talk given  at the XXII Krak\'ow Methodological Conference ``Emergence of the Classical''. Cracow,  October 2018. }} \\
\end{center}

\begin{center} by \end{center}

\begin{center}
H.  Arod\'z\footnote{Email: henryk.arodz@uj.edu.pl} \\ \vspace*{0.2cm}Jagiellonian University,  Cracow, Poland 
\end{center}

\vspace*{1cm}
\noindent {\bf Abstract} \\
Historically,  Ehrenfest's theorem (1927) is the first one which shows that classical physics can  emerge from quantum physics as a kind of approximation. We recall the theorem in its original form. Next, we highlight  its generalizations to the relativistic Dirac particle,  and to a particle with spin and izospin.  We argue that apparent classicality of the macroscopic world can probably be explained within the framework of standard quantum mechanics.

\vspace*{0.5cm}

\section{Introduction}

The principal aim of both  classical and quantum mechanics is to describe  motions of certain physical objects. Both  theories  can be very successfully applied to various  physical objects,  but the sets of these objects  do not coincide.  As  is well known,  classical mechanics gives wrong predictions when applied to microscopic objects such as atoms.  On the other hand,  it seems that quantum mechanics is capable to correctly describe motions  of the  elementary particles as well as   motions of macroscopic bodies, hence it  has a wider  range of applicability. Nevertheless, there are phenomena  description of which requires a theory still more general than quantum mechanics. For example, scattering of elementary particles can lead to creation or annihilation of particles -- here quantum field theory is needed.   Such a generalization of quantum mechanics to quantum field theory is well-known  since the middle of 20th century. There are still some problems with it, but the  prevailing  opinion is that  they concern more its mathematical side than  foundations.  Another departure from standard  quantum mechanics seems to be necessary  when an elementary particle interacts with a very complex, perhaps even  randomly fluctuating or unstable, environment.  Understanding of this  case is rather poor.   To a certain degree  the situation   is analogous to  the well known partition of classical electrodynamics into the theory of the electromagnetic field in vacuum and the electrodynamics of 
continua with  constitutive relations and other additional ingredients.  An effective quantum mechanics in continua is still 
under construction.

 It turns out that  classical mechanics can be derived from quantum mechanics as a kind of approximate theory. Such derivations are usually  called  classical limit of quantum mechanics.  There exist several of them, including the discussed below Ehrenfest type classical limit,  which dates back to 1927 \cite{ehr}, and is  likely  the oldest one.    Its main feature is that it links  solutions of the pertinent fundamental evolution equations, which are  the Schroedinger wave equation in quantum mechanics and the  Newton equation of motion in classical mechanics.  Other kinds of classical limits are carried out on more advanced levels of theory. For instance,  one may derive from quantum mechanics the classical  Hamilton-Jacobi  equation \cite{WKB},   the  Lagrange formalism  \cite{pathint}, or distributions on phase space, see, for example,  \cite{siegel},  \cite{wignerdistr}.

Our main goal here is to recall the seminal paper \cite{ehr},  and to show, using modern examples, how fruitful is the invented by Paul Ehrenfest method for deriving classical mechanics from quantum mechanics. It leads to very interesting extended versions  of classical mechanics featuring, e. g.,  a non relativistic particle with  spin and izospin, or a relativistic particle with spin,   which all emerge from quantum mechanics. Furthermore,  Ehrenfest's theorem  provides a  tantalizing suggestion that  perhaps whole classical physics can be recovered as certain approximations to quantum theories.  Considering wave packets,  we find some  arguments that corroborate this idea.

The present article is addressed to  audience  wider than just  theoretical physicists.   Nevertheless, certain familiarity with basic equations of classical and quantum mechanics is assumed.

The plan of the article is as follows.  First,  we briefly discuss  description of states of a particle  in quantum mechanics with emphasis on the so called wave packets.  Section 3 
is devoted to  the original form of   Ehrenfest's theorem.    In Section 4 we sketch a solution of the main problem with the Ehrenfest method: the lack of relativistic  covariance.  
Next, in the 5th Section  we discuss certain extension of that theorem, which leads to a less known example of classical mechanics  of a point-like particle with spin and  izospin.  Section 6 contains  remarks   on applicability  of quantum mechanics to macroscopic bodies, including a new argument  for practical nonexistence of so called Schroedinger's cats.

\section{ Quantum states of a particle  }

Classical mechanics and quantum mechanics address the same issue:  description of motions of a set of particles.  Such set could consists of just one particle, or a finite number of them. The restriction to finite number of particles is important, because otherwise one would have to use a field theory which is regarded as  different from mechanics for  several important reasons. Classical and quantum mechanics  are structurally similar  to each other  in the sense that  in both theories we introduce a space of states of the particle   and we postulate an equation of motion. They differ in the form of equation of motion:    in classical mechanics  this can be, for example, the  Newton equation, while in quantum mechanics the Schroedinger equation.   Also
 the  spaces of states  are very different. For instance,  for  the simplest single,  point-like particle it is six dimensional phase space in classical mechanics, and  infinite dimensional Hilbert space  in quantum mechanics.   The different  equations of motion,  and different  sets of measurable properties (called observables) for the same set of particles are possible because the spaces of states in classical and quantum mechanics are not identical. Therefore,  we regard  this latter difference as the most important one. 

In  this article we consider the simplest particles, which we describe as  elementary. Particles which possess constituents, for example, hadrons, nuclei, or atoms, are excluded. Physical incarnations of the elementary particles are, e. g., electrons, photons, quarks, or the Higgs particle. 

The  term `point-like particle` used above  is well  justified only in classical mechanics. It refers to the fact that in the simplest case  the state of a single particle at fixed time $t$ is given  by the position and  velocity of the particle.  The position  is represented by a point in the $R^3$ space.  In quantum mechanics the complete description of the state of the particle at a given time $t$ is provided by a smooth wave function $\psi(\vec{x}, t)$   defined  on the $R^3$ space \footnote{For simplicity, we consider here  only so called pure states,  omitting more general mixed states.}.   There is no reason to relate such a quantum particle with a material point  moving in the space. Rather, it should  better be pictured as  a cloud of matter of a very special kind, which is present at all points where the wave function does not vanish.  In particular,  it does not have any  constant shape  or size.   The most peculiar feature of the elementary quantum particle is that it can not be destroyed or created in parts in spite of its spatial extension, while, for example, a drop of water can be divided into parts, and one part evaporated without disturbing the remaining parts.  Physical  processes always involve whole elementary quantum particles,  which are single indivisible entities,  albeit  spatially extended \footnote{In literature this feature is often referred to as the unitarity.}.  With such  picture of the quantum particle, the often discussed and experimentally verified nonlocality of quantum mechanics is natural and rather obvious feature. We shall return to the question what is the best intuitive picture of the quantum particle  in the last Section. 

Certain special  clouds of quantum matter  are called  wave packets.  Roughly speaking, a wave packet is  compact and it consists of  a single bit,  as opposed to  more general quantum states  of the particle which, for example,  can consist of several non-overlapping  compact bits. 
Change in time of any state is described by the Schr\"odinger equation.
It turns out that in the case of particle in empty space typical wave packet  expands.  For example, the width $l(t)$ of a
 three dimensional  (spherical Gaussian) wave packet for a particle at rest is given by the formula \cite{qmtext}
\[ l(t) = \sqrt{l^2_0 + \frac{\hbar^2 t^2}{ m^2 l_0^2}},     \]
where $l_0$  is the initial width  at $t=0$, $m$ is the mass of the particle, $\hbar$   is the Planck constant. This formula implies that the velocity of the expansion monotonically increases to the asymptotic value 
\[ v_0 = \frac{\hbar}{ m l_0}.\]

 The value of Planck's constant is
$ \hbar = 1.0545 \cdot 10^{-27} g \frac{cm^2}{sec}$ , and the masses of electron and proton are, respectively,    $m_e= 9.1 \cdot 10^{-28} g, \;\;\; m_p = 1.67 \cdot 10^{-24} g $.  We would like to draw attention of the reader to the exceedingly small values of these masses. The hydrogen atom  $H$ -- one proton plus one electron, and the hydrogen molecule $H_2$ -- two hydrogen atoms,  also are very very light. If we would like to have  hand-picked one milligram of hydrogen gas \footnote{About 11 ccm  at $0^{o}$C and the normal pressure.}, adding one molecule $H_2$ per second, it would take about   $10^{13}$ years, while the estimated age of our Universe is about $1.4 \cdot 10^{10}$ years.  One should be very cautious when extrapolating our picture of  macroscopic particles to such tiny objects.

It is instructive to compute the asymptotic velocity $v_0$ for various masses and initial widths. 
Let us  first take as the initial width $l_0 = 10^{-8} cm $, which is  the typical atomic size. Then, for the electron we find 
 $v_0 \approx 1160  \: \frac{km}{sec}.$  For a nucleus with the mass   $m= 100 m_p$,  $v_0 \approx 6.4   \: \frac{m}{sec}.$  However, already for  a `speck of dust'  of  size  $l_0=10^{-6} cm$ and mass 
 $m=10^{-4} g$ the velocity is $v_0 \approx 10^{-13}  \: \frac{cm}{sec} \approx 3.2 \cdot 10^{-10} \frac{cm}{year}.$ This means that the wave packet will increase by 1\% during 30 years.  For a drop of water in a fog, $l_0 = 10^{-1} cm$,  $m=10^{-2}g,$ and  $v_0\approx 3\cdot 10^{-17} \frac{cm}{year}$. Thus we see that the electron  in empty space expands  very rapidly.  On the other hand,  the  size of the wave packet for the `speck of dust',  and also  for larger and heavier particles at rest,  remains  practically constant -- the wave packet of  appears as a  `frozen'   blob of quantum matter.

What happens to the wave packet when  we switch on interactions of our quantum particle with other particles?  P. Ehrenfest considered relatively simple case  when the interaction is  described  by a smooth potential $V(\bf{x})$ of a fixed form  (thus he neglected backreaction  of the particle on the other particles).   He proved  the theorem which quite often is summarized by saying  that  in such circumstances the wave packet  moves in the space along a trajectory $\bf{x}(t)$ which obeys Newton's equation of motion
\begin{equation} \ddot{\bf{x}}(t) =  -  \nabla \:V(\bf{x}(t)).     \end{equation} 
Strictly speaking, the actual  content of the theorem is a bit weaker.  Nevertheless, classical equations of the form (1) can be obtained from the theorem after some  further steps.

Our notation is as follows. The dot denotes the derivative with respect to the time $t$. 
The boldface denotes three-dimensional vectors, for
example $\ddot{\bf{x}} = (\ddot{x}^1,  \ddot{x}^2, \ddot{x}^3 )$, where   $x^1, x^2, x^3 $ are Cartesian coordinates in the space, and ${\bf x} = (x^1, x^2, x^3)$.  $\nabla$  is the vector  composed of derivatives with respect to the coordinates, i.e.,  $\nabla = (\partial_1, \partial_2, \partial_3)$, where $\partial_1 = \partial/\partial x^1$, etc., and $\nabla \:V = (\partial_1 V, \partial_2 V, \partial_3 V).$  Summation over repeated indices is understood irrespectively of the level of indices.   ${\bf a}{\bf b} = a^i b^i$ denotes the scalar product of the three-dimensional vectors ${\bf a}= (a^1, a^2, a^3)$ and $ {\bf b}= (b^1, b^2, b^3)$.

\section{ The original form of Ehrenfest's  theorem}

The seminal paper \cite{ehr} is entitled ``Bemerkung \"uber die angen\"aherte G\"ultigkeit der klassischen Mechanik innerhalb der Quantenmechanik''. It  counts merely two and half  pages including the title,  abstract and references. In its first half  the Schroedinger equation for the wave function $\Psi$ is quoted  \footnote{In the present paragraph I copy the original notation from \cite{ehr} in which no special symbol is used for the three dimensional vectors.  The Abstract  in \cite{ehr} clearly indicates that the three dimensional case is considered. In particular,  $\partial/ \partial x $   above should be identified with $\nabla$,  and  $\partial^2/ \partial x^2 $  with the Laplacian $\Delta$.}, 
\[ - \frac{\hbar^2}{2m} \frac{\partial^2 \Psi}{\partial x^2} + V(x)  \Psi = i \hbar \frac{\partial \Psi}{\partial t}, \]
as well as its complex conjugation.   
 Next it is stated that these equations imply the following relations 
\begin{equation} \frac{d Q}{dt} = \frac{1}{m} P, \;\;\;   m \frac{d^2 Q}{dt^2} = \frac{d P}{dt}= \int \! dx \:\Psi \Psi^* (- \frac{\partial V}{\partial x}),  \end{equation}
where \[Q(t) \equiv  \int\! dx  \: x  \Psi \Psi^*,  \;\;\;\mbox{and}  \;\;\;\;   P(t)  \equiv i \hbar \int \! dx \:\Psi \frac {\partial  \Psi^*    }{\partial x}. \]
Details of the derivation are omitted,   except for the remark that the second relation in formulas (2)  is obtained  with the help of integration by  parts.  P. Ehrenfest assumes that the spatial extension of the wave packet is small compared with macroscopic distances 
(nota bene, he uses the name `wave packet'  for the product $\Psi \Psi^*$).  

Commenting on his results,  P. Ehrenfest  underlines similarity of the second  relation in (2)  to Newton's equation of classical mechanics. He is satisfied  with such qualitative correspondence,  and does not attempt to make it more precise.  In particular,  he does not even mention the  approximation
\[  \int \! dx \:\Psi \Psi^* (- \frac{\partial V}{\partial x}) \approx - \frac{\partial V(Q)}{\partial Q}, \]
probably because he  knew that it would be a hard task to formulate it in a rigorous manner. In fact, this approximation is  the subject of numerous  nontrivial investigations till nowadays.  Only with this approximation the second relation (2)  turns into Newton's equation (1) 
if we identify $ Q(t)$  with  ${\bf x}(t).$

The second part of the paper has the subtitle `Bemerkungen'. It is devoted to the one dimensional Gaussian wave packet for a free particle ($V=0$).  Its  explicit form is presented, and the spreading out discussed.  The paper ends with the observation  that in the case of a very heavy particle  the Gaussian wave packet expands very slowly,  while for  proton very rapidly.  

Paper \cite{ehr} is  very important, indeed, for  at least two reasons. First,  P. Ehrenfest  has shown that quantum mechanics does not  contradict classical mechanics, but rather generalizes it -- the latter can be regarded as a very good approximation to the former  for a large class of physical phenomena. 
Second, he  pioneered  derivations of various kinds of classical equations of motion from underlying quantum mechanical models.  Two important examples of this kind are outlined below.

\section{ Lorentz covariant formulation of the Ehrenfest method }

There is a problem with Lorentz covariance in the Ehrenfest approach to classical limit:  because  the standard expectation values do not have  clear relativistic transformation law,  the classical mechanics derived from Lorentz covariant quantum mechanics based  on,  e.g., the Dirac equation, is not covariant. Hence, it can hardly be accepted as the correct classical limit. This problem is explicitely pointed out  in  \cite{HilWou}.   

It turns out that there exists a modification of the Ehrenfest method which yields  Lorentz covariant result \cite{aro1}.  Below we give a  description of the results. Our main point here is that there is no single classical mechanics that follows from the underlying quantum theory.  Instead, we obtain an infinite sequence of classical theories, which approximate the quantum theory with better and better accuracy and, unfortunately, with a  complexity  rapidly increasing to the level that renders such  classical  theories impractical.

This  paragraph contains certain technical details given here for the readers interested in the theoretical physics aspects of the work \cite{aro1} --  it can be omitted by not interested ones.    In the improved approach, we start  from a new definition of expectation values, which respects the Lorentz covariance. In this definition, the integral over the three Cartesian coordinates $x^1, x^2, x^3$ is replaced by an integral over 
three  new spatial coordinates  in a special  coordinate system in the Minkowski space-time. In this system, the time axis is replaced by 
a time-like line $X^{\mu}(s)$ in the Minkowski space-time.  This line will ultimately coincide with  the  classical  trajectory  associated with the wave packet. The three  new spatial coordinates parameterize the directions perpendicular to this line (in the Minkowski sense).
The Cartesian time coordinate $t$ is replaced  by the proper time coordinate  $s$ on  that line. Next, the Dirac equation  is transformed to these new coordinates.  The evolution parameter is not the laboratory time $t$, but the proper time $s$.  There are certain consistency conditions for the new expectation values which result from the requirement that     
the line $X^{\mu}(s)$  stays close to the  wave packet,  which evolves according to the Dirac equation. The explicit form of the wave packet is not needed. The consistency conditions  imply the  classical 
equations of motion  for  $X^{\mu}(s)$, and for other quantities like spin.   Their form is  approximate one  in the sense that  all terms proportional to $1/m^2$  or to higher powers of $1/m$ have been neglected. This is justified  because $m$ is assumed  to have a  large value.    We use the Foldy-Wouthuysen transformation.

The starting point -- the Dirac equation for a single electron -- has the form 
\[  \gamma^{\mu} \left(\frac{\partial}{\partial x^{\mu}} + i  A_{\mu}\right)\psi  + i m \: \psi =0.  \]
It replaces  the Schroedinger equation  considered by P. Ehrenfest.  
 $A_{\mu}(x)$  in the Dirac equation denotes the so called four-potential of  electromagnetic field. It encodes information about electric and magnetic fields in which the  electron moves.  By assumption,  it does not include the field generated by  the considered electron.  Furthermore,  we assume that the mass $m$ is large, in accordance with the discussion of spreading out of  wave packets in Section 2.  For convenience, we use the notation in which the Planck constant $\hbar$ and the velocity of light in vacuum $c$ are not visible  --  as if  $c=\hbar =1$  (the notation commonly referred to as `the natural units').  
We also assume that the particle has unit electric charge. Summation over repeated indices is understood.  We use the standard relativistic four dimensional notation as explained in, e.g., \cite{relat}.

The modified Ehrenfest method yields the classical equations of motion which read:   
\[\hspace*{-3cm} m \ddot{X}_{\mu} =  F_{\mu\nu}\dot{X}^{\nu}+ \frac{1}{2m} \epsilon_{\nu\lambda\sigma\alpha}\dot{X}^{\lambda} (\delta^{\beta}_{\mu}-  \dot{X}^{\beta} \dot{X}_{\mu}) \:W^{\sigma} \:\partial_{\beta}F^{\alpha\nu} \] 

\vspace*{-0.4cm}
\begin{equation} \hspace*{2.0cm} + \frac{1}{2m} (\delta^{\sigma}_{\mu}-  \dot{X}^{\sigma} \dot{X}_{\mu}) \:C^{\rho\nu}\: \partial_{\rho}F_{\nu\sigma}   + \frac{1}{2m}  \dot{X}^{\rho} \dot{X}_{\nu} \: C_{\mu\sigma}\:\partial_{\rho} F^{\nu\sigma},     \end{equation}

\vspace*{-0.3cm}
\[ \frac{dW^{\lambda}}{ds} =  - \dot{X}^{\lambda} \ddot{X}_{\mu}W^{\mu} + \frac{1}{m} (\delta^{\lambda}_{\mu}-  \dot{X}^{\lambda} \dot{X}_{\mu}) \:W_{\nu}\:F^{\mu\nu}  \hspace*{2cm} \] 

\vspace*{-0.4cm}
 \begin{equation} \hspace*{2.5cm} + \frac{1}{m} (\delta^{\lambda}_{\mu}-  \dot{X}^{\lambda} \dot{X}_{\mu})\:Z_{\sigma \rho}\: F^{\mu\sigma,\rho}  + \frac{1}{m} (\ddot{X}^{\lambda} P^{\nu}_{\nu} +  \ddot{X}_{\nu}P^{\nu\lambda}).     \end{equation}
Technical details again:   the dot denotes the derivative $d/ ds$,  where  $s$ is the proper time  along the classical trajectory $X^{\mu}(s)$.  The proper time  replaces the time $t$   present in Eqs.\ (2).  Furthermore, $\partial_{\mu}$ stands for the partial derivative  $\partial/ \partial x^{\mu}$, and 
$F_{\mu\nu} = \partial_{\mu}  A_{\nu} - \partial_{\nu} A_{\mu} $  is the electromagnetic field strength  tensor.  It is composed of the electric and magnetic fields.  $\epsilon_{\nu\lambda \sigma \alpha}$ (the  so called totally antisymmetric  symbol)  is equal to $0, 1, -1$ depending on the values of the Greek indices, for instance, $\epsilon_{0123} =1$.  
The spin four-vector  $ W^{\sigma}$   is related to  the expectation value of the quantum spin operator.  In the particular case of constant electric and magnetic fields,  equations  (3) and (4) reduce to the well-known Bargmann-Michel-Telegdi equations  for  a relativistic  particle with spin.

The relativistic classical equation of motion for a point-like  particle with the unit electric charge ($e=1$) in the external electromagnetic field   that is  usually  given in textbooks has the form 
\begin{equation}  m \ddot{X}_{\mu} =  F_{\mu\nu}\dot{X}^{\nu}.   \end{equation}
It precedes the quantum mechanics and also the concept of spin. We see  that 
it is a small part of  equation (3) above. Moreover,  equation (5)  does not take into account the spin of the particle, which in equations (3), (4)  is represented by $W^{\mu}$.  In many important tasks,  for example, in  calculations of trajectories  of  electrons or protons in accelerators,  one has to use equations which take into account the spin in order to achieve the desired accuracy  -- equation (5) is not good enough.   In practice,  certain simplified version of  equations  obtained  with the Ehrenfest method  is used. Such  nontrivial and successful applications corroborate   the correctness of  the attitude that classical equations of motion should be derived from underlying quantum theory.  On the other hand,  there are many problems in which the spin is not important. In such cases the old equation (5)  gives very good predictions for trajectories of the particle.

The classical  variables    $ C^{\rho\nu}(s), \;Z_{\rho\sigma}(s),   \;P^{\nu\lambda}(s)$  are related to entanglement of quantum observables: position with momentum, position with spin,  and momentum with spin, respectively,  \cite{aro1}.  In principle,  also equations of motion for  $ C^{\rho\nu}(s), \;Z_{\rho\sigma}(s)$ and $P^{\nu\lambda}(s)$ are needed for the mathematical completeness of the system of equations.  They can be obtained with the help of the (modified) Ehrenfest method, but in  practice one usually eliminates these variables by making certain simplifying assumptions. For example, in most situations all terms in the second line  of equation (3), and also in the second line of (4),  can be omitted.  Then we do not need equations of motion for    $ C^{\rho\nu}(s), \;Z_{\rho\sigma}(s),   \;P^{\nu\lambda}(s)$.     If the  equations  of motion  for these classical  variables were included, one would get even more accurate 
classical approximation to the quantum mechanics, but at the price of having to deal with a much larger set of equations.

\section{ Classical mechanics of a point-like particle with spin and  color }

This example of derivation of classical  mechanics  is  interesting  because   prior to the pertinent quantum theory such a classical theory had not been known at all.  Once derived, it has turned out to be  a useful tool for theoretical investigations of  quark matter.  Quarks have  special charges,  called  color and  weak izospin,  which make them  sensitive  to the so called non-Abelian gauge fields. Both the non-Abelian gauge  fields and the quarks  as constituents  of the material world  were discovered  in 1960's and 70's.  Certain particular
version of the non-Abelian gauge field is called the Yang-Mills field. Below we outline the basic features of  the classical limit for a quantum particle that interacts with the Yang-Mills field.  The resulting classical theory  describes motion  of a point particle, known as  the particle with color or izospin,   in  certain fixed Yang-Mills field.

Historically,  the first attempt to derive  classical equations of motion for a point particle    interacting with the  Yang-Mills field was made by S. K. Wong in 1970  \cite{wong}.  
 Classical state of this  particle  at given time $t$ is represented  jointly by:  the so called classical izospin vector  $I^a(t)$, where the index $a$  takes values 1, 2 and 3;   the  position $\mathbf{x}(t)$;  and the velocity $\dot{\mathbf{x}}(t)$.   The derivation  given by Wong  does not use the Ehrenfest method.  For that matter,   it should rather be  described as an educated guess  based on symmetry  principles and algebraic structure of the Dirac equation. In consequence, his  equations respect the Lorentz invariance as well as  the so called  gauge invariance, but they miss  the spin of the particle and  certain less obvious classical  variable, as explained below.  We will not present here these equations in order to avoid overloading this article with technical details. Interested reader may consult  the original paper by Wong  or \cite{aro2}.

More systematic derivation is based on  the Ehrenfest method \cite{aro2}.   We  consider expectation values  of   the following quantum observables:  the position $\hat{\mathbf{x}}$, the so called kinetic momentum $ \hat{\mathbf{p}} - \mathbf{A}^a \hat{T}^a$,  the spin $\hat{\mathbf{S}}= (\hat{S}^i)$, the izospin $\hat{\vec{T}}= (\hat{T}^a)$, and the product of the spin and izospin operators $\hat{J}^{ai} =  \hat{T}^a \hat{S}^i$. The hat $\hat{}$ means that these objects are operators in  pertinent Hilbert space.  The indices $a, \:i $ take values 1,2, and 3.  The three vectors  $\mathbf{A}^a$  represent the Yang-Mills field.  They are counterparts of the  electromagnetic  vector potential  $\mathbf{A},$ which is a part of the four-potential $A_{\mu}$  introduced in the previous Section,  $(A_{\mu}) = ( A_0, \:\mathbf{A})$.

Expectation values of these observables become the classical dynamical variables.  Furthermore, the pertinent  quantum evolution equation yields the counterpart of the Newton equation and a few other equations.  In this manner we  obtain the classical mechanics with the following  classical variables that characterize the point particle with izospin:    the position  $\mathbf{x}(t)$,  the velocity $\dot{\mathbf{x}}(t)$,   the classical  izospin  $I^a(t)$,   the classical spin vector  $\mathbf{S}(t)$, and a novel classical variable 
 $ J^{ai}(t)$. 
 
The novel dynamical variable $J^{ai}(t)$ is the expectation value of the operator $\hat{J}^{ai} $.  It can be regarded as the  three vectors $\mathbf{J}^a(t)$, $a=1,2,3$, with their components enumerated by the index $i$.  In spite of the fact that the operator $\hat{J}^{ai}$ is the product of operators 
$\hat{T}^a$ and $\hat{S}^i$,  its expectation value does not have to be  equal to the product  $I^a(t) S^i(t)$,  because in general  expectation value of  product of operators is not equal to  product of expectation values of the operators.  

The Ehrenfest method  not only reveals the new classical variable --  it also  shows that there are relations, traditionally called constraints,   between the classical variables, which reflect the fact that the classical variables are defined as the expectation values in the same quantum state $\psi({\bf x}, t).$ These constraints have the following form
\[  4  J^{i a}  S^i = I^a,  \;\;\;   4 J^{ia}  J^{ib} = (\frac{1}{4} - \mathbf{S}^2) \: \delta_{ab} +   I^a I^b.  \]

To summarize,   applying the Ehrenfest method we have discovered that   Wong's equations of motion for the classical point particle with izospin  are rather oversimplified version of the more adequate equations.  In particular, we have found the new classical variable $J^{ai}(t)$, which appears because the particle possesses both spin and izospin.

\section{Conclusion and  remarks} 

{\bf 1.}   Ehrenfest’s theorem and its generalizations show that  classical mechanics of particles can be reinterpreted in terms of expectation values, with pertinent quantum states being the wave packets.  In this way,  the relation between classical and quantum mechanics, viewed as the relation  between old and new theories, acquires  the perfect form:   the new theory is more general and more accurate, and it rather encompasses the old one instead of contradicting it in all respects.  Furthermore,  the method  used by Ehrenfest -- the emphasis on properties and  evolution of expectation values -- has turned out to be very fruitful as the tool for improving existing classical  theories. In Section 4  we have seen such improvement  in the case of classical particle in electromagnetic field. The method can also provide completely new classical mechanics,  unknown prior to quantum theory,  as discussed in  Section 5 on  the example of particle with spin and izospin.

{\bf 2.} The  enormous success of the Ehrenfest method suggests that perhaps no  part of  the material world  is purely classical,  that  quantum mechanics embraces all physical phenomena \footnote{With possible exception for gravitational phenomena. So far there is no  experimental evidence for  quantum nature of gravitation.}, and that the classical world is fictitious in the sense that it exists only as  certain theoretical approximation to the real world  \footnote{Here we touch the philosophical problem to what extent it really does not exist.  Interesting philosophical analysis of a related problem  can be found in \cite{heller}.}. 
  Such assumption of  absolute quantumness of the seemingly classical macroscopic world leads to the following question: why we do not see in nature  isolated macroscopic bodies in typical quantum states  such as, e. g.,  wave packets spatially extended over sizable distances  (in literature dubbed  `Schroedinger's cats').  To explain their absence, one can propose  a new theory which deviates from quantum mechanics in the macroscopic world, and essentially coincides  with it  in the micro-world. The recently popular   Continuous Spontaneous Localization theory \cite{CSL} is of this kind. 
 One should also mention 
the decoherence  phenomenon \cite{zeh},  \cite{zurek}, in which  states of a quantum system are very quickly transformed into the so called  mixed states, due to strong  interactions with environment. Here the absence of widely extended wave packets of macroscopic particles is explained by the presence of interactions with an environment. Which mixed state (`pointer state’) appears  at the end of the process of decoherence  of a concrete wave packet still is the matter of  many investigations. It is a difficult problem, and there are many related hypotheses, some with picturesque names, e.g.,  `quantum Darwinism' \cite{zurek2}.   The decoherence phenomenon belongs to the realm of  effective quantum mechanics in continua, mentioned in the Introduction.

The author prefers another viewpoint: we think that one can provide an  explanation for the apparent absence  of quantum phenomena in the macroscopic world using the standard quantum mechanics. An interesting  possibility is that such extended quantum states of heavy isolated particles are possible in principle, but that they are hardly achievable in reality. The main difficulty is  that a spatially extended state has to be produced as such,  because  wave packets of  very heavy particles practically do not expand.  This can be rather difficult task. For illustration,  let us  consider the following thought experiment.  Suppose that we  can produce a kind of hydrogen-like `atom' in which the electron is replaced  by a heavy (in comparison with electron)  particle  of the mass $M= 10^{-6} g$, and the proton with an  even more massive particle. Next, let us excite it in order to increase its spatial size.  Highly excited states close to ionization  threshold have a macroscopic size --  there is no theoretical upper bound  on the size of excited atoms.  Finally, we ionize that `atom' -- this would provide the heavy particle (`electron') in an extended  quantum state of the size of the `atom'.  The trouble is that the energy needed for the ionization is of the order  $10^{13}$ GeV, as  a simple calculation shows, while the highest  achievable at present  energies of particles are of the order $10^{4}$ GeV only. 

Another thought experiment involves quantum harmonic oscillator. This system is  ubiquitous in physics -- it arises as a very good approximation to many complex systems. Classical harmonic oscillator consists of a particle of mass $M$ subject to  a force which increases proportionally to the distance from a fixed point, called the center, to the particle. The strength  of the force is characterized  by a constant $k$.  Quantum theory of such object predicts that the least energy state  has the form of  a wave packet of the size  $l = \sqrt{2 \hbar/ \sqrt{M k}}$. Now, let us  take the particle roughly  of the size of a droplet of water from a fog.  Its  radius is $r= 10^{-1} cm$ and the mass  $M= 10^{-2} g$.  We are interested in situations such that  $l$ is much larger than  $r$ -- then the wave packet will be much larger than the classical radius of the particle. Simple calculation  shows that  the constant $k$ has to be exceedingly small, namely $k \ll 4\cdot 10^{-48} g/sec^2$.  Sizable force appears only when the distance from the center is of the order $10^{40} cm$. Let us recall that the light year counts about $10^{18} cm$.  Construction of such a  feeble  harmonic oscillator  is far beyond the present day engineering. On the other hand, if we take a more realistic value $k =1 \:g/sec^2$,  the condition $ l \gg r $ is satisfied only if  $M \ll 4 \cdot 10^{-50} g$ -- the  mass  incomparably smaller than the mass of electron. Such particle certainly is not macroscopic.

{\bf 3.}  Let us return to the question from Section 2:  what is the best intuitive picture of  elementary quantum particle. Such a picture can be very helpful if it is adequate, or very misleading  when wrong. In our opinion,  many mysteries, controversies, and so called paradoxes that are discussed in literature on quantum mechanics   arise in a large part from inadequate images of the quantum particle. As we have written in Section 2, we prefer to regard
 the quantum particle as a cloud of quantum matter. Its main feature is that it can be created or annihilated  as a whole  -- it is impossible to  have one half of electron.  Notwithstanding our views,   we  admit that  there exist other pictures as well. It seems that the most popular one is that   actually there exists exactly point-like material particle which has a concrete position in space  at each time, but we do not know that position. What is known is  merely the probability of finding this point-like  particle in a chosen volume of the space. It is calculated as the integral of the modulus squared of the wave function over that volume.  We think that by adopting such  image of the quantum particle  one simply carries over to quantum mechanics the picture from classical mechanics \footnote{This is precisely what was done in the prequantum theory of atoms with the Bohr-Sommerfeld  rules. This theory is in fact classical one. The Bohr-Sommerfeld rules serve only as a tool for selecting particular classical trajectories.}. This can not be justified,  especially if we regard classical mechanics of point-like particles as a secondary theory which is derived from quantum mechanics.  Therefore we  should base our intuitions solely on the Schroedinger equation, and on the actual  mathematical representation of the states of the particle as wave functions,  forgetting completely about the classical mechanics.   

The picture of point-like quantum particle with  concrete yet unknown location in the space  may be motivated also by unjustified enhancement of the probabilistic interpretation of quantum mechanics. It is known for sure from  numerous experiments that outcomes of  measurements are distributed with certain probability, which can be calculated with the help of quantum mechanics if we assume the so called Born rule. The point is that there is no experimental evidence for the probabilistic character of quantum mechanics without invoking an experiment. Thus, we  may suppose that it is a specific coupling  between the two systems:  the quantum particle and a very special physical macroscopic apparatus -- the measuring apparatus -- that is responsible for the probabilistic nature of outcomes of experiments. We adhere precisely to this view.  

 To summarize, we prefer the picture of elementary particle as a cloud of quantum matter. The probabilistic outcomes of measurements  are due to  interaction of the particle  with a  macroscopic  measuring apparatus. For us, such views are quite natural  corollaries to Ehrenfest's theorem.

\section{Acknowledgement}
I would like to thank the Organizers of the XXII Krak\'ow Methodological Conference  for the invitation to give the talk  at this very   pleasant and stimulating  meeting.

\end{document}